\documentclass{PoS}

\usepackage{amsmath}
\usepackage{bm}

\usepackage{graphicx}

\newcommand{\be}{\begin{equation}}
\newcommand{\ee}{\end{equation}}

\newcommand{\ka}{\kappa}

\newcommand{\beq}[1] {\begin{equation}\label{#1} }
\newcommand{\eeq} {\end{equation} }
\newcommand{\bea}[1]{\begin{eqnarray}\label{#1} }
\newcommand{\eea}{\end{eqnarray}}

\newcommand{\si}{\sigma}

\def\beqn{\begin{eqnarray}}
\def\eeqn{\end{eqnarray}}

\def\beq{\begin{equation}}
\def\eeq{\end{equation}}
\def\bea{\begin{equation}}
\def\eea{\end{equation}}

\def\al{\alpha}
\def\bt{\beta}
\def\Ga{\Gamma}

\def\ka{\kappa}
\def\si{\sigma}

\def\te{\theta}

\def\lam{\lambda}

\def\om{\omega}
\def\ep{\epsilon}

\def\l{\left (}
\def\r{\right )}

\def\fr{\frac}
\def\la{\label}
\def\hs{\hspace}
\def\vs{\vspace}

\def\ran{\rangle}
\def\lan{\langle}
\def\ov{\overline}

\title{Recent Progress in SUSY GUTs}

\ShortTitle{SUSY GUTs}

\author{\speaker{K.S. Babu}\thanks{Work is supported in part by the US Department of Energy Grants
DE-FG02-04ER41306 and DE-FG02-ER46140;  OSU-HEP-11-05.}\\
        Department of Physics, Oklahoma State University, Stillwater, OK 74078, USA\\
        E-mail: \email{babu@okstate.edu}}

\abstract{After a brief review of the motivations for grand unification, I discuss the main
challenges facing realistic SUSY GUT model building.  Achieving doublet--triplet splitting without
fine--tuning is chief among them.  Symmetry breaking should occur consistently
without unwanted Goldston bosons, $\mu$ term of order TeV for the MSSM Higgs fileds should emerge
naturally, and realistic fermion masses with small quark mixing angles and large lepton mixing angles should be
generated with some predictivity.  Significant progress has been made over the years
towards achieving these goals in the context of supersymmetric $SO(10)$ GUT.  A complete
$SO(10)$ model is presented along this line wherein, somewhat surprisingly, 
the GUT scale threshold corrections to the gauge couplings are found to be small.
This results in a predictive scenario for proton lifetime.  An interesting correlation between the
$d=6$ ($p\to e^+\pi^0$) and $d=5$ ($p\to \ov{\nu }K^+$) decay amplitudes is observed.
This class of models predicts that both proton decay modes should be observable with an improvement
in the current sensitivity by about a factor of five to ten.}

\FullConference{35th International Conference of High Energy Physics - ICHEP2010,\\
		July 22-28, 2010\\
		Paris France}

\begin{document}
\section*{}
\vspace*{-0.5in}

Motivations for unifying the strong, weak and electromagnetic forces are manyfold \cite{ps,gg,georgi}.  The experimental
observation that electric charges are quantized ($|Q_{\rm proton}| = |Q_{\rm electron}|$ to better than 1 part in $10^{21}$)
has a natural explanation in grand unified theories (GUT) owing to their non--Abelian nature.
The miraculous cancelation of chiral anomalies that occurs among each family of
the SM fermions has a symmetry--based
explanation in GUTs. $SO(10)$ GUT, for example, is automatically free of such anomalies \cite{georgi}.  GUTs provide a natural understanding of the quantum
numbers of quarks and leptons.  This point is worth emphasizing further.  All quarks and leptons of a family,
including the right--handed neutrino ($\nu^c$) needed for generating small neutrino masses via the seesaw mechanism, are organized into
a {\bf 16}--dimensional spinor representation of $SO(10)$, as shown in Table 1.  The gauge symmetry $SO(10)$ contains five independent
internal spins, denoted as $+$ or $-$ signs (for spin--up and spin--down) in Table 1.  Subject to the condition that the number of
down spins must be even, there are 16 combinations, which form the irreducible ${\bf 16}$ dimensional spinor of $SO(10)$.
The first three spins denote color charges, while the last two are weak charges.  There are three independent combinations of color spins,
identified as the color degrees of freedom ($r,b,g$).  Going top down in each column of Table 1, one sees
that in addition, there is a fourth color, identified as lepton number \cite{ps}.  Thus quarks and leptons
are unified under the GUT symmetry.  The first and the third columns (and similarly the second and the fourth) are left--right conjugates.
Thus $SO(10)$ contains Parity as part of the gauge symmetry.  Furthermore, the same {\bf 16} multiplet unifies quarks with anti-quarks, and leptons
with anti-leptons.  In fact, $SO(10)$ symmetry is the maximal gauge symmetry that is chiral with sixteen particles (members of one
family).  Hypercharge (and thus electric charge) of each fermion follows from the formula $Y = \frac{1}{3} \Sigma (C) - \frac{1}{2} \Sigma (W)$,
where  $\Sigma (C)$ is the summation of color spins (first three entries) and $\Sigma (W)$ is the sum of weak spins (last two entries).  Thus
$Y$ for the $e^c$ field is $Y(e^c) = \frac{1}{3} (3) - \frac{1}{2} (-2) = 2$.  Note that $Y$ (and thus $Q$) must be quantized.
Such a simple organization of matter is remarkably beautiful and can be argued as a strong hint for GUTs.  SUSY GUTs have further
empirical support
from the observed unification of gauge couplings at a high energy scale $M_X \approx 10^{16}$ GeV.  In Fig. 1, left panel, we demonstrate this unity
of forces in a fully realistic $SO(10)$ SUSY GUT \cite{bpt}.  Another remarkable feature of $SO(10)$ GUTs is that the small neutrino masses inferred from
neutrino oscillation data suggest the scale of new physics ($\nu^c$ mass scale) to be $M_{\nu^c} \sim 10^{14}$ GeV, which is close to $M_X$.
 $M_{\nu^c}$ is inferred from  the effective neutrino mass operator ${\cal L}_{\rm \nu} = LLH_u H_u/M_{\nu^c}$ ($L$ is the lepton doublet and $H_u$ 
 is the Higgs doublet), using $m_\nu \sim 0.05$ eV and $\left\langle H_u \right\rangle \sim 246$ GeV as inputs.  In a class of $SO(10)$ models discussed further here, $M_{\nu^c} \sim M_X^2/M_{\rm Pl} \sim 10^{14}$ GeV quite naturally \cite{bpw}.  The decay of $\nu^c$ can elegantly explain the observed baryon asymmetry of the universe via leptogenesis.  Finally, as exemplified later, the unification of quarks and leptons into GUT multiplets can be quite powerful in realizing predictive
 frameworks for fermion masses, perhaps in association with flavor symmetries.  

\begin{table}[h]
{\footnotesize
\begin{center}
\begin{tabular}{||c|c|c|c||}\hline\hline
$u_r:~\{-++~+-\}$ & $d_r:~ \{-++~-+\}$ & $u^c_r:~\{+--~++\}$ & $d^c_r:~ \{+--~--\}$ \\
$u_b:~\{+-+~+-\}$ &  $d_b:~ \{+-+~-+\}$ & $u^c_b:~\{-+-~++\}$ &  $d^c_b:~ \{-+-~--\}$ \\
$u_g:~\{++-~+-\}$ &  $d_g:~\{++-~-+\}$ & $u^c_g:~\{--+~++\}$ &  $d^c_g:~ \{--+~--\}$ \\
$~\nu:~\{---~+-\}$ &  $~e:~~ \{---~-+\}$ & $~\nu^c:~\{+++~++\}$ &  $~e^c:~ \{+++~--\}$ \\ \hline\hline
\end{tabular}
\end{center}
\vspace*{-0.2in}
\caption{Quantum numbers of quarks and leptons. The first 3 signs refer to color charge, and the last three to weak charge.
To obtain hypercharge, use $Y = \frac{1}{3}\Sigma(C)-\frac{1}{2}\Sigma (W)$.}
}
\end{table}

While extremely well motivated, constructing fully realistic SUSY GUTs is not so trivial.  Chief among the challenges is
the so--called doublet--triplet (DT) splitting problem.  Even in the simplest of GUTs, based on $SU(5)$ gauge symmetry \cite{gg},
the smallest irreducible representation is 5 dimensional, which contains the SM Higgs doublet.  This means that the
Higgs doublet will be accompanied by a GUT partner, which is a color triplet scalar ($5=2+3$ is the relevant 
math).  This state must have a mass of order
$10^{16}$ GeV, or else it would lead to rapid proton decay.  The DT splitting challenge is to naturally make the doublet
component of the ${\bf 5}$--plet light (of order $10^2$ GeV), while maintaining its color triplet partner superheavy.
In minimal SUSY $SU(5)$, this is done by extreme fine--tuning, of order one part in $10^{14}$.  The relevant superpotential
is $W = \overline{5}_H (\lambda 24_H + M)5_H$, where the $(\overline{5}_H, \,5_H$) contain the MSSM Higgs doublets ($H_d,\,H_u$),
and where $\left\langle 24_H \right\rangle = V. \,{\rm diag}(1,\,1,\,1,\,-3/2,\,-3/2)$ breaks the gauge symmetry down to that of
the SM.  The masses of the color triplet and $SU(2)_L$ doublet fields are then $m_T = \lambda V + M$, $m_D = \lambda V - (3/2)\,M$.
One chooses $\lambda V$ and $M$ to be both of order $10^{16}$ GeV, but with the condition $\lambda V = (3/2)\,M
+ {\cal O} (10^2)$ GeV, so that the doublet remains light.  Such a severe fine--tuning raises questions about the naturalness of the
model.  

In SUSY $SO(10)$ the situation for DT splitting is much better.  The adjoint Higgs $A(45)$ in $SO(10)$ can acquire a vacuum expectation value
(VEV) $\lan A\ran ={\rm i}\si_2\otimes {\rm Diag}\l a, ~a,~a,~0,~0\r$.  The coupling $H(10)A(45)H'(10)$ of two 
$10$--plets  would result in heavy color triplets with massless $SU(2)_L$ doublets -- without fine--tuning \cite{dw}.  There are
a variety of issues that need to be addressed.  Symmetry breaking must be complete without unwanted pseudo--Goldstone bosons,
the VEV of the adjoint should be stable against higher dimensional operators, unification of gauge couplings should be maintained,
and the theory should be consistent with proton
lifetime limits.  All these issues have been successfully addressed recently \cite{bpt} by making use a set of small dimensional
Higgs fields, as shown in Table 2.  An anomalous  ${\cal U}(1)_A$ symmetry of possible string origin and a $Z_2$ symmetry are also used, for
stabilizing the doublet mass.  In Table 2, $k$ is a positive integer, which will be taken to be 5.  
\begin{table}[h]

\vs{-0.5cm}
\label{t:U-Z-charges} $$\begin{array}{|c||c|c|c|c|c|c|c|c|c|c|c|}
\hline
\vs{-0.4cm}
 &  &  &  &  &  &  &  & & &&\\

\vs{-0.5cm}

& ~A(45)~& ~H(10)~&~ H'(10)~& ~C(16) ~& ~ \bar C(\ov{16})~&~Z ~&~ S ~  & ~C'(16) ~ & ~\bar C'(\ov{16}) & 16_{1,2}&16_3\\

&  &  &  &  &  &  &  &  & &&\\

\hline

\vs{-0.5cm}
 &  &  &  &  &  &  &  & & &&\\

~Q~& 0 &1  &-1  &{(k+4)}/{2k}  & -{1}/{2}   &{2}/{k}   & {2}/{k}   & {(k-4)}/{2k}  & -{(k+8)}/{2k}   &q_{1,2}&-{1}/{2}\\

\vs{-0.5cm}
&  &  &  &  &  &  &  &  & &&\\

\hline
\vs{-0.5cm}
 &  &  &  &  &  &  &  & &&&\\
\vs{-0.5cm}
~\om ~& 1 & 0 & 1 & 0 & 0  & 1  & 0  & 0  & 0  &P_{1,2}&0\\

&  &  &  &  &  &  &  &  & &&\\

\hline

\end{array}$$
\vspace*{-0.2in}
\caption{${\cal U}(1)_A$ and $Z_2$ charges $Q_i$ and $\om_i$ of the superfield $\phi_i$.}
\end{table}
\noindent The superpotential consistent
with the symmetries is 
{\small
\begin{eqnarray}
W &=& M_A{\rm tr}A^2+\fr{\lam_A}{M_*}\l {\rm tr}A^2\r^2+\fr{\lam_A'}{M_*} {\rm tr}A^4
+ C\l \fr{a_1}{M_*}ZA+\fr{b_1}{M_*}C\bar C+
c_1S\r \bar C' +C'\l \fr{a_2}{M_*}ZA+\fr{b_2}{M_*}C\bar C+c_2S\r \bar C
\la{sup-ACCbar}\nonumber \\
&+& \lam_1HAH'+\l \lam_{H'}SZ^{k-1}+ {\lam'}_{H'}Z^k\r \fr{(H')^2}{M_*^{k-1}}
+\lam_2H\bar C\bar C +\fr{\lam_3}{M_*}AH'CC'~.
\la{W-H}
\end{eqnarray}
}
This superpotential consistently breaks the $SO(10)$ gauge symmetry down to the SM symmetry in the SUSY limit without generating
unwanted Goldstone bosons.  The first three terms of $W$ induce the VEV for $A$, the fourth and fifth terms guarantee absence of
Goldstones \cite{br}, and the remaining terms achieve doublet--triplet splitting without fine--tuning \cite{dw}.  At the minimum we have $\left \langle C \right
\rangle = \left \langle \bar{C} \right \rangle = c$, $\left \langle C' \right
\rangle = \left \langle \bar{C}' \right \rangle = 0$, with $c$ determined by the Fayet--Iliopoulos term of ${\cal U}(1)_A$ symmetry.  The 
Higgs doublet mass is zero in the SUSY limit to all orders.  Once SYSY breaking is turned on, the VEV of the $A(45)$ no longer has the zeros, which
are modified to be entries of order $m_{\rm SUSY}$.  This in turn induces $\mu$ term of order $m_{\rm SUSY}$ for the MSSM Higgs fields.
In Fig. 1 (left panel), the evolution of the three gauge couplings is displayed, which takes into account the threshold corrections of
the model.  Interestingly, the threshold corrections in the model in the $10+ \bar{10}$ sector (of the $SU(5)$ subgroup) cancel between
the matter fields and the gauge boson fields.  Owing to this cancelation, the model becomes very predictive for proton lifetime. We find
a correlation between the $d=6$ gauge boson mediated $p \rightarrow e^+ \pi^0$ and the $d=5$ Higgsino--mediated $p \rightarrow \overline{\nu} K^+$
decay amplitudes: $M_{\rm eff}\simeq 10^{19}{\rm GeV} \cdot \l \fr{10^{16}{\rm GeV}}{M_X}\r^3
 \l \fr{1/100}{r}\r \l \fr{3}{\tan \bt }\r $.  $M_X$ controls $ p \rightarrow e^+ \pi^0$, while $M_{\rm eff}$ controls
 $p \rightarrow \overline{\nu}K^+$.  This is plotted in Fig. 1 (right panel) for varying $r=M_\Sigma/M_X$ ($\Sigma$ is a color
 octet Higgs field).  Also plotted are the current experimental limits from these decays. One concludes that both modes
 should be observable with an improved sensitivity of about five to ten.  

\begin{figure}
\centering
    \includegraphics[scale=0.35]{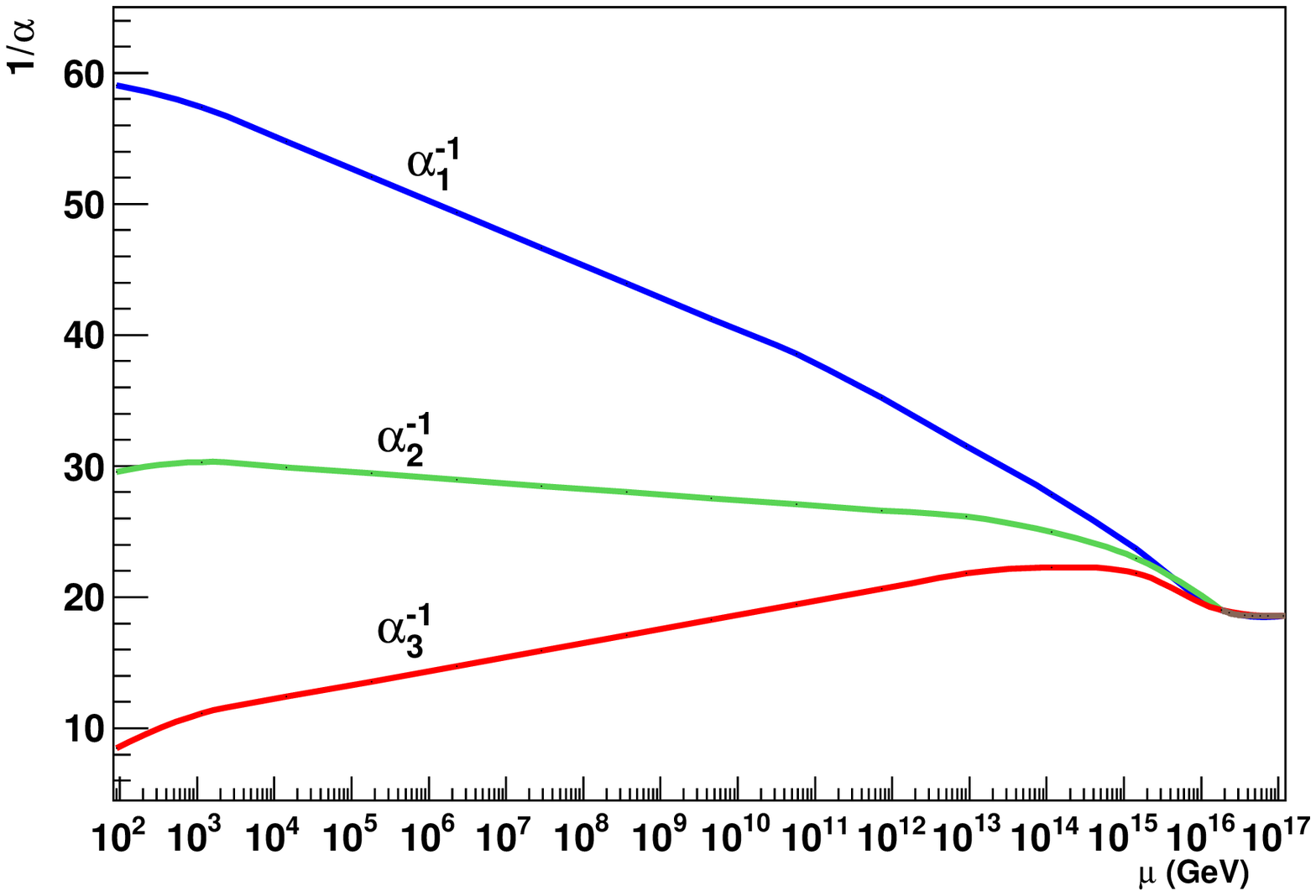}
    \hspace{0.5cm}
    \includegraphics[scale=0.37]{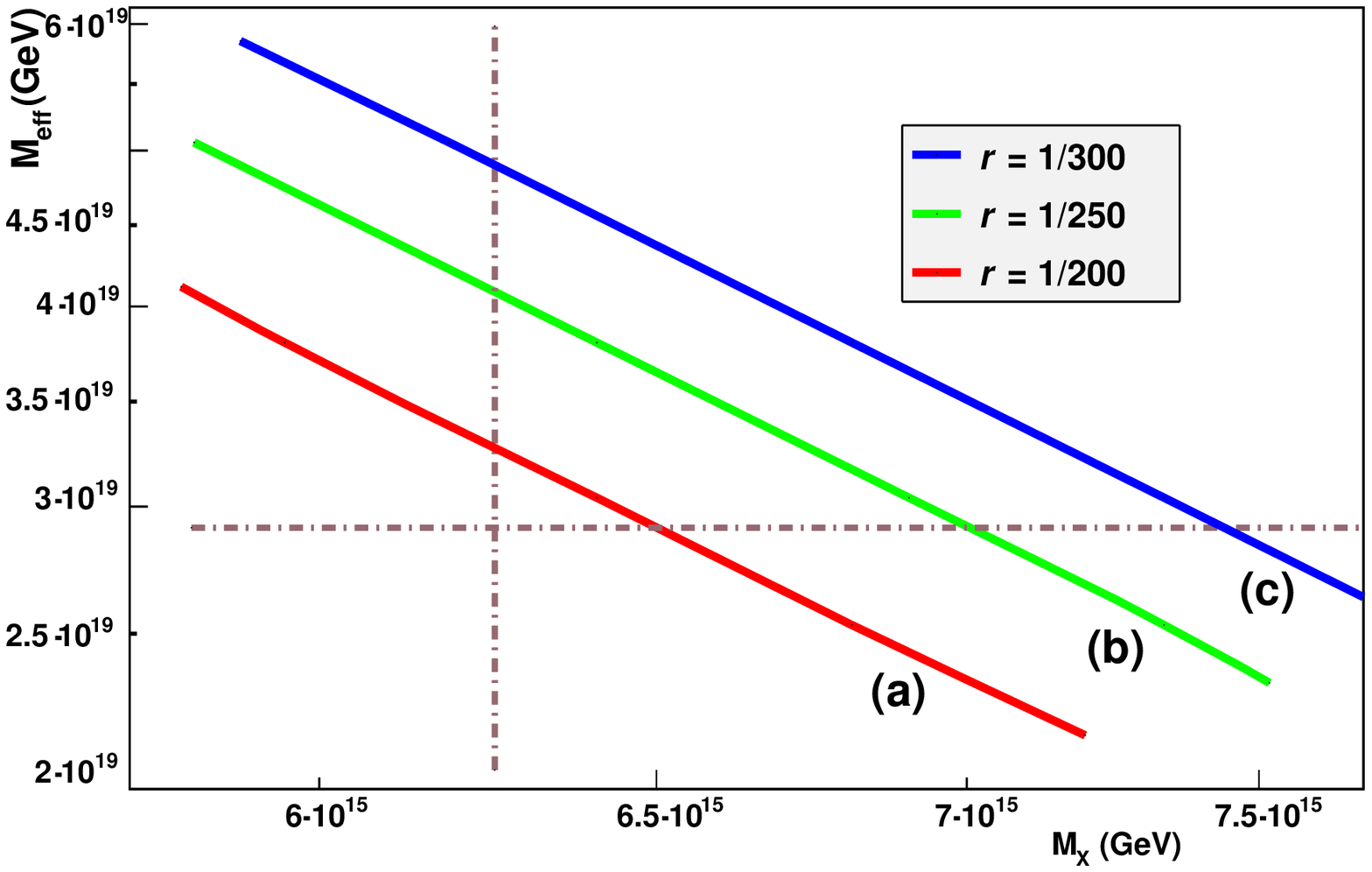}
    \caption{Left panel: Gauge coupling evolution including threshold corrections.  Right panel:
    Correlations between $M_{\rm eff}$ and $M_X$ for $m_{\tilde{q}} = 1.5$ TeV, $m_{\tilde{W}} = 130$ GeV,
and $\al_3(M_Z)= 0.1176$. {\bf (a)}: $r=1/200$. {\bf (b)}: $r=1/250$. {\bf (c)}: $r=1/300$. The vertical
and horizontal dashed lines correspond to the experimentally allowed lowest values of $M_X$ and $M_{\rm eff}$
which arise from limits on $\Ga^{-1}(p\to e^+\pi^0)$ and $\Ga^{-1}(p\to \bar{\nu }K^+)$.}
\end{figure}

Realistic and predictive fermion masses and mixings can be obtained within this framework by assuming a flavor $Q_4$ symmetry, under
which the first two families of $16$ form a doublet.  The mass matrices for up and down quarks, charged leptons and Majorana
neutrinos have the form \cite{bpt,bpw}:
\begin{eqnarray}
M_u = m_U^0\left(\begin{matrix}
0 & \epsilon' & 0 \cr -\epsilon' & 0 & \sigma \cr 0 & \sigma & 1
\end{matrix}\right)\hs{-0.1cm},\,M_{d,e} = m_D^0\left(\begin{matrix}
0 & \kappa_{d,e}\epsilon'+\eta' & 0 \cr -\kappa_{d,e}\epsilon'-\eta' & \kappa_{d,e}\xi_{22}^d & \sigma+\kappa_{d,e}\epsilon \cr 0 & \sigma +
\kappa_{d,e}\bar{\epsilon}& 1
\end{matrix}\right)\hs{-0.1cm},\,M_R = m_R^0\left(\begin{matrix}
b & 0 & 0 \cr 0 & b & a \cr 0 & a & 1
\end{matrix}\right)\,\,\,
\end{eqnarray}
where $\ka_{d}=1$ and $\ka_e=3$.   There are fewer parameters than observables in this setup, which results in predictions. A consistent
fit for all masses and mixing parameters is obtained with the choice
$\si =0.0508, \ep =-0.0188+0.0333i, \ov{\ep }=0.106+0.0754i ,
\ep'=1.56\cdot 10^{-4}, \eta'=-0.00474+0.00177i,
\xi_{22}^d=0.014 e^{4.1i}$ at the GUT scale.  Along with central values of charged lepton masses, we obtain for the quarks,
$m_u(2~{\rm GeV})=3.55~{\rm MeV},~m_c(m_c)=1.15~{\rm GeV},\,
m_d(2~{\rm GeV})=6.45~{\rm MeV}$, $m_s(2~{\rm GeV})=137.6 {\rm MeV},~m_b(m_b) = 4.67~{\rm GeV}$.
For the CKM mixings we obtain:
$|V_{us}|=0.225$, $|V_{cb}|=0.0414~,|V_{ub}|=0.0034~,|V_{td}|=0.00878~,
\ov{\eta }=0.334~,\ov{\rho }=0.12$, and thereby $\sin 2\bt =0.663$. All these are in a good agreement with experiments.
For neutrinos, the Dirac mass matrix is obtained from $M_u$ by replacing $\ep'\to -3\ep'$.
With $\te_{12}\simeq 30^o$ and $\te_{23}\simeq 43^o$ as inputs, we obtain $m_2/m_3\simeq 0.13$
and $\te_{13}\simeq 3.6^o$ as predictions.
Such a fit is realized by choosing $a=0.0252e^{-0.018i}$, $b=1.61\cdot 10^{-6}e^{-1.592i}$,
and $M_0=1.89\cdot 10^{13}$~GeV.  One sees broad, although not precise, agreement with
data.  The model succeeds in obtaining large neutrino mixings along with small quark mixings.



\end{document}